# Amorphous Photonic Lattices:
# Band Gaps, Effective Mass and Suppressed Transport

**Mikael Rechtsman[1], Alexander Szameit[1], Felix Dreisow[2], Matthias Heinrich[2], Robert Keil[2], Stefan Nolte[2], and Mordechai Segev[1]**

[1]*Physics Department and Solid State Institute, Technion, 32000 Haifa, Israel*
[2]*Institute of Applied Physics, Friedrich-Schiller-Universität Jena, Max-Wien-Platz 1, 07743 Jena, Germany*

**Abstract**

We present, theoretically and experimentally, amorphous photonic lattices exhibiting a band-gap yet completely lacking Bragg diffraction: 2D waveguides distributed randomly according to a liquid-like model responsible for the absence of Bragg peaks as opposed to ordered lattices containing disorder, which always exhibit Bragg peaks. In amorphous lattices the bands are comprised of localized states, but we find that defect states residing in the gap are more localized than the Anderson localization length. Finally, we show how the concept of effective mass carries over to amorphous lattices.

Conventional intuition holds that, for a solid to have an electronic band-gap, it must be periodic, allowing the use of Bloch's theorem. Indeed, the free-electron approximation implies that Bragg scattering in periodic potentials is precondition for the formation of a band-gap. But this is obviously untrue: looking through a window reveals that glassy silica ($SiO_2$), although possessing no long-range order, still displays an electronic band-gap spanning the entire frequency range of visible light, which is responsible for the lack of absorption. Amorphous band gaps were put on firm theoretical ground in 1971 [1], and experiments in electronic systems followed [2,3]. With the major progress in photonic crystals [4,5], it is natural to explore amorphous photonic media. Indeed, these were studied theoretically (e.g. [6-8]), and microwave experiments demonstrated the existence of a band-gap [6,9]. However, amorphous band-gap media have never been studied experimentally in the optical regime.

Here, we present the first experimental study of amorphous photonic lattices exhibiting a band-gap: a liquid-like two-dimensional (2D) array of randomly-organized evanescently-coupled waveguides. We show that the bands in this medium are separated by gaps in its spatial spectrum, despite a total lack of Bragg scattering. We find that the liquid-like distribution plays a key role in the absence of Bragg peaks, because in a periodic lattice containing disorder, Bragg peaks are always present. The bands in amorphous photonic media are comprised of inherently localized Anderson states, yet we find that these lattices support strongly localized defect states, whose widths are much narrower than the Anderson localization length. Finally, we show the existence of a region of negative effective mass (anomalous diffraction), and how to observe it through transport experiments. Amorphous photonic lattices are a test-bed for the properties of general amorphous systems, because optics offers the possibility to precisely engineer the potential, as well as to employ nonlinearity under controlled conditions. Hence, this work paves the way for unraveling features that are much harder to access experimentally in other systems.

The evolution of the optical wave $\Psi(x,y,z)$ in our paraxial system is described by

$$i\bar{\lambda}\frac{\partial \Psi(x,y,z)}{\partial z} = -\left[\frac{\bar{\lambda}^2}{2n_0}\nabla_\perp^2 + \Delta n(x,y)\right]\Psi(x,y,z) \tag{1}$$

where $\bar{\lambda} = \lambda/2\pi$ is the reduced wavelength, $n_0$ is the refractive index of the bulk (fused silica, in our experiments), and $\Delta n(x,y)$ is the refractive index modulation caused by the random distribution of identical waveguides. Equation 1 is similar to the Schrödinger equation, when $\bar{\lambda}$ is replaced by $\hbar$, $n_0$ by the mass $m$, $\Delta n(x,y)$ by the potential $-V(x,y)$, and the spatial coordinate $z$ by time $t$ [10]. This similarity was used in many recent experiments, demonstrating concepts from solid-state physics using optical settings [10]. One important example is Anderson localization, which was realized in photonic lattices [11] using the transverse localization scheme [12]. To obtain the band structure, one needs to solve Eq. 1, by substituting $\Psi(x,y,z)=\varphi(x,y)e^{i\beta z}$ with the propagation constant $\beta$. The resulting eigenvalue equation cannot be solved by applying Bloch's theorem, since $\Delta n(x,y)$ is not periodic. Rather, as known from Anderson localization in 2D, the eigenmodes are fully localized functions ("Anderson states"). Consequently, one must solve using the full refractive index profile to find the eigenvalue spectrum.

Our liquid-like amorphous lattices are made of individual waveguides with no long-range order. They are therefore fundamentally different from quasicrystals, which possess long-range order and hence do exhibit pronounced Bragg diffraction [13,14]. We emphasize that it is impossible to create an amorphous medium only by introducing uncorrelated disorder to a periodic structure, in which case the Bragg peaks remain. Rather, as we show below, a liquid-like 2D (or 3D) structure lacks any diffraction peaks. We distribute the waveguides as a snapshot of atoms in a liquid, by the use of the Metropolis Monte Carlo method with a repulsive interatomic force [15]. Such structures result in random $\Delta n(x,y)$, with one particular realization shown in Fig. 1(a). Clearly, the modulus of the Fourier transform of $\Delta n(x,y)$ has no delta-function peaks [Fig. 1(b)], implying diffuse scattering. The broad elliptic rings are similar to those seen in x-ray scattering on liquids. To highlight the fundamental difference between the amorphous liquid-like structure and a periodic lattice containing disorder, we study the latter under the same parameters. Figure 1(c) shows a disordered pattern where the position of each waveguide is randomly and independently perturbed, from a perfect square lattice. In sharp contrast to the amorphous case, the modulus of the Fourier transform of this pattern *displays pronounced Bragg peaks* [Fig. 1(d)]. The presence of long-range order in the perturbed crystal always gives rise to Bragg scattering, exemplifying the fundamental difference between amorphous media and a crystal containing disorder.

We fabricate a waveguide array corresponding to the structure of Figs. 1(a,b) using femtosecond direct laser writing [16]. Figure 2(a) depicts a microscope image of the structure. All waveguides have identical structure (slightly elliptic, due to fabrication constraints), and the refractive index step defining them is $\Delta n = 9 \times 10^{-4}$. As Fig. 2(b) shows, the Fourier transform of this amorphous structure displays a total lack of Bragg peaks. We calculate the eigenmodes of this structure, and show in Fig. 2c the values of $\beta$ for $\lambda=633$nm. Figure 2c reveals a sizeable gap in the spatial spectrum, despite a total lack of periodicity (and lack of Bragg scattering) in Fig. 2b. This defies a common argument [17], stating that gaps open at the boundary of the Brillouin zone because the degeneracy of states there is broken by the periodicity of the potential. This argument is based on a perturbation theory where the potential is weak, which is inapplicable for our system because the potential is comparable to the kinetic energy.

A nice feature offered by optical settings described by Eq. 1 is the ability to test the properties of the system by tuning parameters independently, with the most notable one being $\lambda$, which can be tuned continuously. The inset in Fig. 2c reveals that the gap width is exponentially decreasing with $\lambda$ (when all other parameters are fixed), until it closes at 900nm. This can understood from Eq. 1: increasing $\lambda$ leads to a larger "kinetic energy" (transverse Laplacian), and thus to relative weakening of the potential inducing the gap. As shown below, this wavelength dependence of the gap provides an efficient tool for exploring the properties of amorphous photonic media.

The band-gap in amorphous photonic systems calls for some intuition. In the amorphous medium of Figs. 1 and 2, all waveguides are identical, but their spacing is random. It is instructive to plot the size of the gap as a function of the variance of the inter-waveguide spacing (Corresponding to a variance in the tight-binding hopping parameter). We plot that in Fig. 2d for $\lambda$=633nm, and find that there is a sizeable gap as long as the normalized standard deviation is below 18%. The key feature of our liquid-like structure giving rise to the large band-gap is the short-range order: the waveguides have similar spacing, thus they form bonds of similar energy and in turn, populate the bands and leave the gap empty. This explanation holds for bands arising from guided modes of the individual waveguide as well as for higher bands arising from unbound states, as in the single-mode waveguides structure of Fig. 2.

The realization of a photonic band gap in our amorphous structure merits discussion. There have been many suggestions of structures that yield a photonic band gap: traditional photonic crystals where the refractive index is periodic in one, two [18], and three dimensions [4], to Coupled Resonator Optical Waveguides (CROWs) [19], and even anti-resonant devices [20]. All of these structures are periodic and so they must involve Bragg scattering. Band-gaps in amorphous media therefore represent a fundamentally different paradigm.

To visualize the gap experimentally, we introduce a defect waveguide in the structure: a single waveguide with a refractive index maximum lower by $\Delta n_d=4.5\times10^{-4}$ than all other waveguides. Calculations show that this results in a single, negative, defect state residing directly in the band-gap (Fig. 3a), for $\lambda=633nm$. By contrast, at $\lambda=875nm$ the gap is extremely small; hence the defect state occurs where the bands merge (Fig. 3b). Consequently, when we launch a $\lambda=633nm$ beam into the defect waveguide, the beam stays strongly confined throughout propagation, because it excites a highly localized defect state. This is shown experimentally in Fig. 3(c) depicting the intensity structure of the beam exiting the amorphous lattice. The coupling to all nearby waveguides is greatly suppressed, in spite of their close proximity, as light is guided in a defect state residing in a sizeable band-gap. In contrast, at $\lambda=875nm$ there is no gap, hence a 875nm beam launched into the same waveguide tunnels to many other waveguides (Fig. 3d). Thus, by demonstrating the presence of the defect state, we have experimentally proved the existence of a band-gap in this amorphous optical system.

We now demonstrate that a defect-state in an amorphous system is much more localized than the Anderson localization length. The potential in our system is 2D (x and y), thus all states are inherently localized with any amount of disorder [21], i.e., the bands solely contain localized

Anderson states. It is therefore interesting compare Anderson localization in our amorphous lattice with the light confinement in a defect state residing within the gap. **This question is of fundamental importance, and has never been addressed experimentally.** Figure 4 shows the results: a defect state (residing in the gap) is much more localized than the Anderson localization length. The experiments on Anderson localization require ensemble averaging [11], where one has to average over multiple realizations of disorder to obtain meaningful results. Figures 4(a),4(b) depict the intensity profile of light trapped in a defect mode (Fig. 4(a)), and light that is Anderson-localized (Fig. 4(b)). The defect state is invariant under averaging, in contrast to a wave-packet made of Anderson modes (which are part of the band), where a single realization does not reveal the arrest of transport by virtue of disorder. Figures 4c,4d show the cross-section taken through the experimental (4a,4b) and simulated results under the experimental conditions. These figures reveal that the defect state is much narrower than the localization length. This is because the defect state exhibits an isolated eigenvalue, with no other states of similar energy with which it may hybridize and thus delocalize. On the other hand, although the other eigenstates are themselves localized, they reside in similar environments, thus hybridize with one another and "spread out". Interestingly, we find that the width of a defect state residing in the band-gap of our amorphous structure is comparable to that of a defect state in the gap of a fully periodic crystal of the same (mean) lattice spacing.

Next, we show that the concept of *effective mass* at the band edges carries over from the periodic to the amorphous case. In photonic lattices, the effective mass is defined as the inverse of the second derivative of the propagation constant with respect to the transverse momentum [10,22]. Hence, the effective mass in photonic lattices is measured by varying the transverse

Bloch momentum of a launch beam and observing the variation of its transverse group velocity [22], relying on having Bloch modes. However, in amorphous systems the dispersion relation is discontinuous, the Anderson states are all localized with zero transverse velocity ($\Delta n$ is z-invariant), hence one cannot use this method. Instead, we quantify the effective mass directly through Newton's second law: introducing a known variation of the potential (varying much slower than the spacing between waveguides), launch a wavepacket, and observe its trajectory. An example is shown in Fig. 5, where we add a weak, slowly varying, sinusoidal function to the structure $\Delta n(x,y) \rightarrow \Delta n(x,y) + \alpha \sin(2\pi x/L)$, where $L$ is the width of the sample, and *(0,0)* is taken to be at the center [23]. Then, we construct a beam from a superposition of eigenstates within a small range of $\beta$ at close vicinity (such that effective mass can be defined), and launch it near the center coordinate at the input facet. With $\alpha>0$ the force acts in the +x direction. Hence, for a positive effective mass, the beam would be deflected towards +x, whereas for a negative effective mass it would propagate in the –x direction, opposite to the force direction. The amount of deflection is inversely proportional to the effective mass (at lowest order approximation). We demonstrate this concept through simulations. Figure 5a shows the average deflection of a wavepacket composed of eigenstates at the top of the first band (indicated in the inset). It moves in the +x direction, demonstrating positive effective mass. Figures 5b,5c show the wavepacket at the input and output facets, indicating the deflection to the right. Here, a Gaussian envelope was imposed on the initial wavepacket, and the results are ensemble-averaged over 100 realizations of the amorphous pattern. Figures 5d-f, and 5g-k are analogous to Figs. 5a-c but for the cases of the bottom of the first band (where we observe negative effective mass), and the top of the second band (where the effective mass is positive). Clearly, the concept of effective mass carries over to amorphous systems. This suggests interesting experiments, especially in the nonlinear

regime – where an attractive nonlinearity would act the opposite for positive and negative effective mass [24], even though the potential is completely random. Unfortunately, demonstrating effective mass through transport requires beam shaping beyond our current experimental reach.

In this Letter, we demonstrated the first experiments on band gaps in amorphous photonic structures, and showed that they provide an excellent test-bed to explore the universal features of systems with no long-range order. We found defect states embedded in the band gap of the amorphous medium, which are fundamentally different from the Anderson states comprising the bands in amorphous media. Such defect modes can guide light very efficiently through random media. Finally, we studied the concept of effective mass in amorphous optical systems. These findings raise many intriguing questions, such as, how does nonlinearity impact light evolution in such media? Would a repulsive nonlinearity lead to suppression of transport for wave-packets with negative effective mass? The issue of nonlinear waves in amorphous media is very exciting. Would nonlinear phenomena, such as spontaneous pattern formation, exist in amorphous media? Do such media support shock waves? Is it possible to generate solitons in amorphous media? If so, then how would solitons move through the random potential and interact with one another, as waves or as particles? Our setting will help exploring these concepts, and understand the true universal nature of amorphous media.

**Figure Captions:**

**Fig. 1.**

Amorphous photonic lattices: theoretical results. (a) Refractive index profile, $n(x,y)$, of an amorphous waveguide structure. The index varies from *1.45* to *1.45+9×10$^{-4}$*. (b) Modulus of the Fourier transform of this structure, which shows no Bragg peaks. (c) $n(x,y)$ for a square lattice with superimposed uncorrelated disorder. (d) Modulus of the Fourier transform of (c), displaying clear Bragg peaks. Note that (b) and (d) were averaged over many different realizations to highlight the Bragg peaks.

**Fig. 2:**

Amorphous photonic lattices: experimental results. (a) Microscope image of the input facet of the waveguide lattice. (b) Modulus of Fourier transform of (a), showing no Bragg peaks. (c) Eigenvalue spectrum of amorphous waveguide lattice at $\lambda$=633nm, showing a band gap. Inset: size of the band gap as a function of $\lambda$. (d) Variation of band gap size as a function of standard deviation of the spacing between waveguides.

**Fig. 3:**

(a),(b) Eigenvalue spectrum of the amorphous lattice with a defect waveguide (refractive index 1.45+4.5×10$^{-4}$) at $\lambda$=633nm and $\lambda$=875nm, respectively: (a) contains a defect mode in its gap, while (b) has no gap and therefore no defect mode. (c),(d) Experimental images of the light intensity at output facet, showing a highly localized defect mode in the gap in (c), but not in (d).

**Fig. 4:**

(a) Experimental ensemble-averaged (over 10 samples) output beam for light launched into the defect waveguide. Ensemble-averaging does not affect the shape of the defect mode. (b) Experimental ensemble-average output beam for *633nm* light input on 30 different non-defect waveguides in different local environments of the disordered pattern. The ensemble-average wavepacket exhibits Anderson localization. (c) Semilog light intensity cross-sections for the experiments of (a) and (b). (d) Numerical results (ensemble-averaged over 100 realizations of disorder) for curves corresponding to those in (c) derived from beam-propagation simulations. The defect state is always narrower than the Anderson localization length.

**Fig. 5:**

Quantifying the effective mass through deflection. Top, middle, bottom rows: deflection of a wave-packet taken from the top edge of the first band, bottom edge of first band, and top edge of the second band, respectively. Left, middle, and right column: deflection vs. propagation distance, input light distribution, and output light distribution, respectively. The deflections in (c),(f),(k) indicate positive, negative, and positive effective mass, respectively. All results are ensemble averaged over 100 realizations of the amorphous pattern.

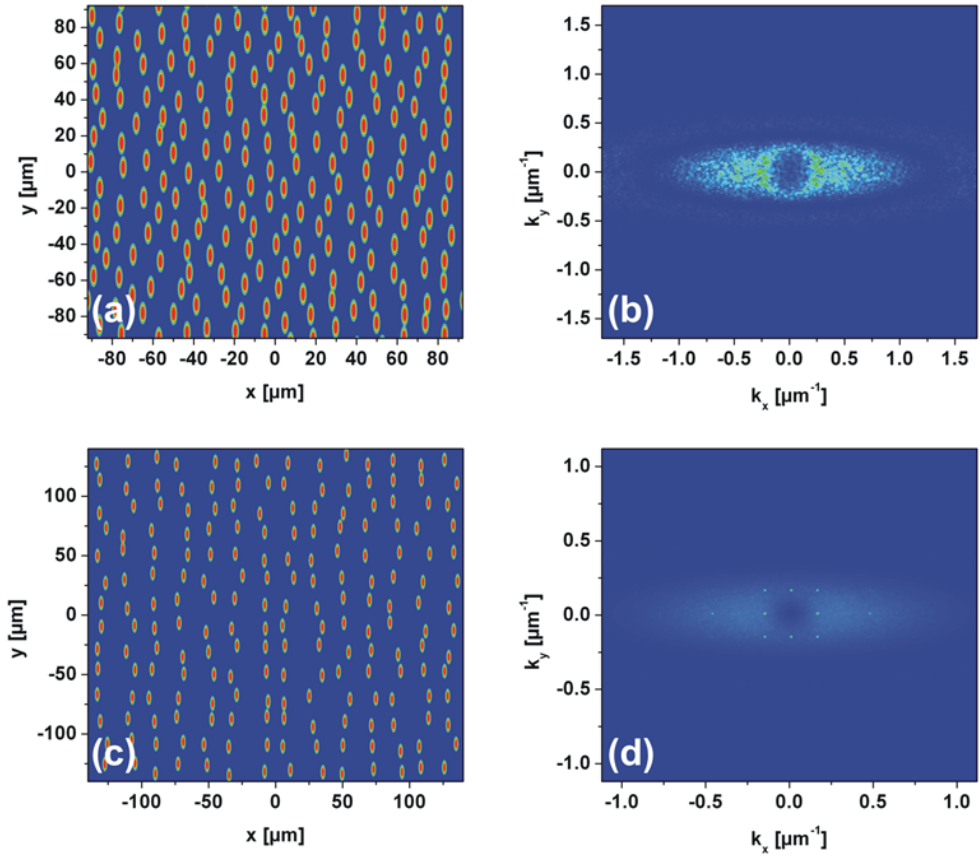

Figure 1

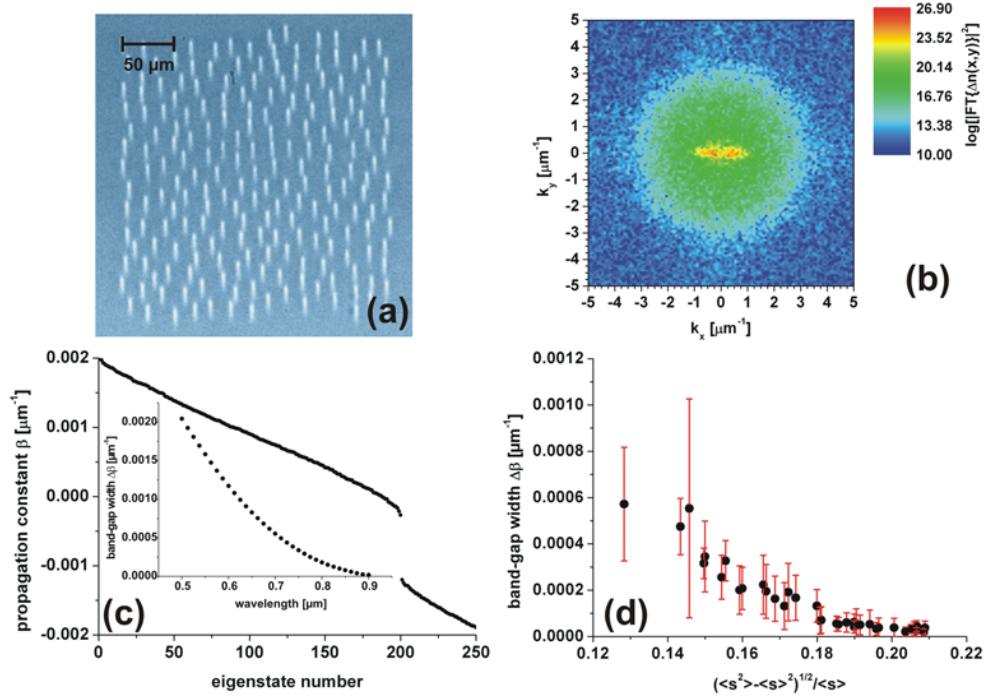

Figure 2

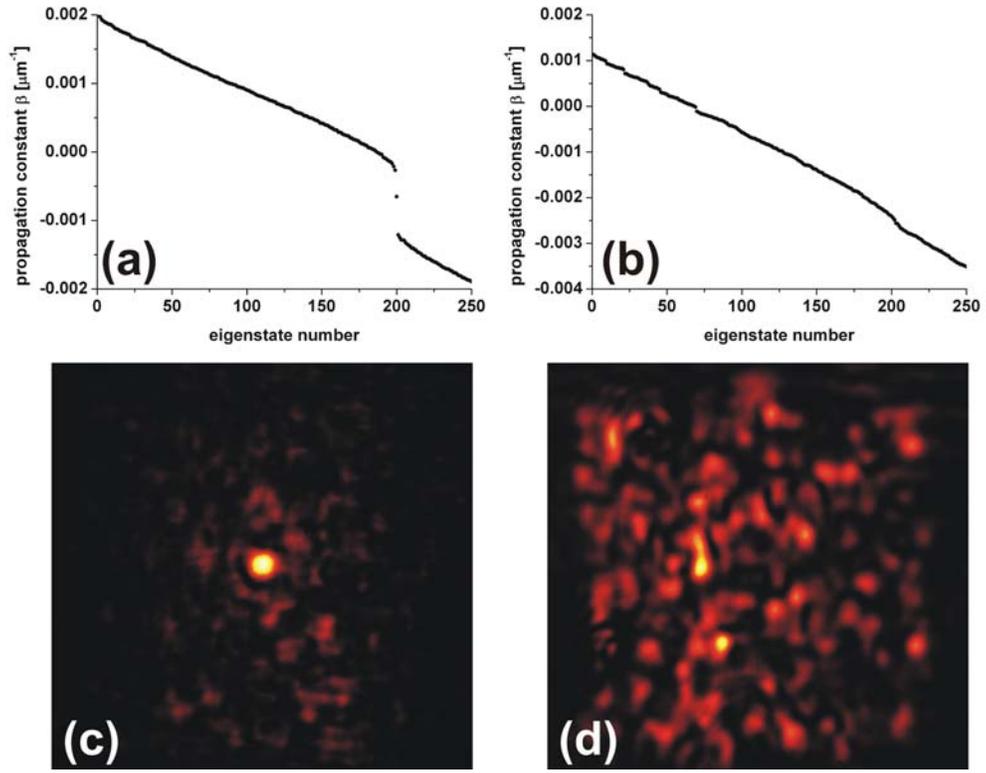

Figure 3

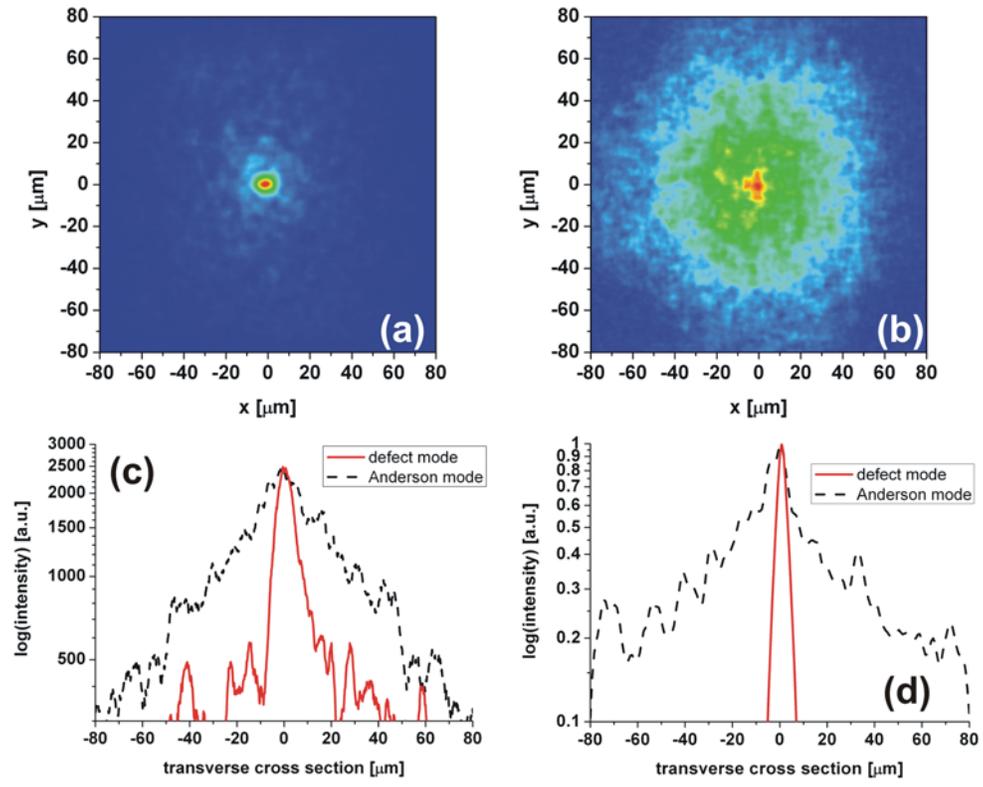

Figure 4

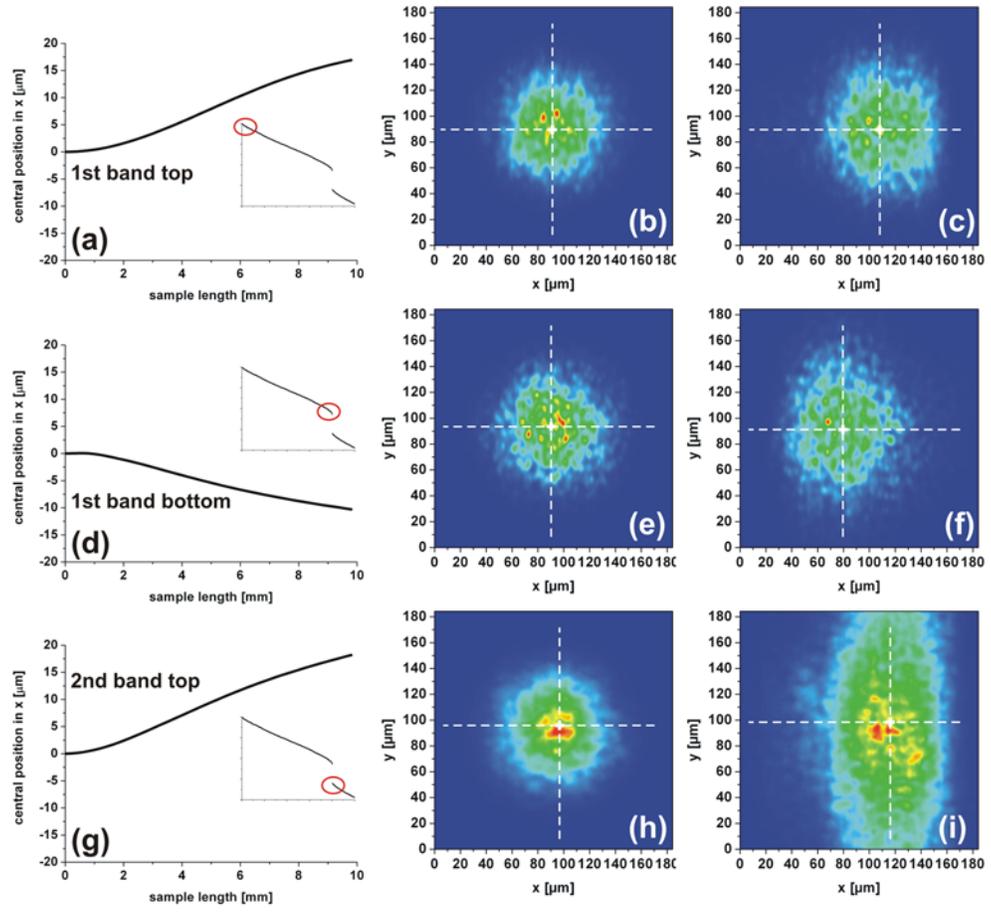

Figure 5

# Amorphous Photonic Lattices:
# Band-Gaps, Effective Mass and Suppressed Transport

Mikael Rechtsman[1], Alexander Szameit[1], Felix Dreisow[2], Matthias Heinrich[2], Robert Keil[2], Stefan Nolte[2], and Mordechai Segev[1]

[1]*Physics Department and Solid State Institute, Technion, 32000 Haifa, Israel*
[2]*Institute of Applied Physics, Friedrich Schiller University Jena, Max-Wien-Platz 1, 07743 Jena, Germany*

## Table of Contents

## A. Calculating the band structure of amorphous photonic lattices

The paper presents several calculations of the band structure of amorphous photonic lattices. These figures display the propagation constants (eigenvalues) of the eigenmodes of Eq. 1, which describes the propagation of optical waves, under the paraxial approximation, in such lattices. The eigenmodes, or the stationary solutions of Eq. 1, are all localized states (Anderson states) and the propagation constants associated with them form bands separated by gaps, in a fashion similar to periodic lattices. The propagation constants and the Anderson modes are calculated using the plane-wave expansion method [5].

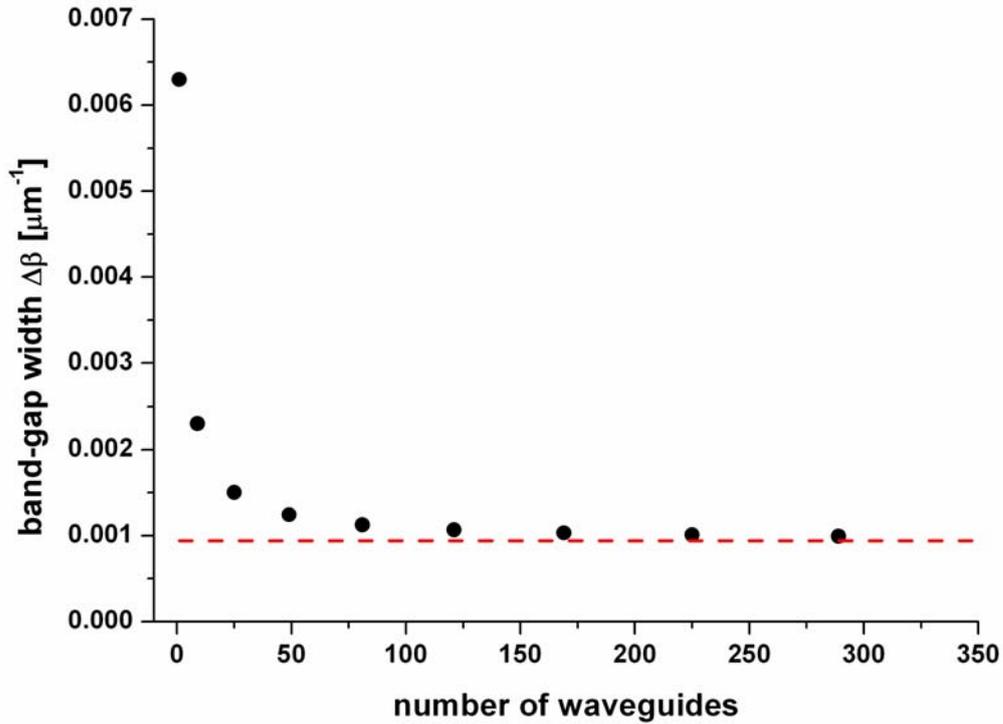

**Fig. S1:** Band-gap of a square lattice of waveguides as a function of number of waveguides in the system. A plane-wave expansion method is used to calculate the band structure, and hence the band-gap, for systems composed of $N^2$ waveguides, where $N=1,3,5,7,9,11,13,15,17$, and 19. Only odd integers are used because when $N$ is even the symmetry of the square lattice yields the exact band-gap for periodic boundary conditions. As can be seen from the figure, 200 waveguides is enough to achieve convergence of the band gap. The lattice spacing between the waveguides is *14 μm*.

In the paper, band structures are displayed for systems with and without defect and at different optical wavelengths. In these calculations, we employ periodic boundary conditions on a system composed of 200 waveguides. In order to prove that taking 200 waveguides is sufficient to achieve convergence of the band structure, we examine a square lattice of waveguides where the band structure can be calculated to high accuracy using Bloch's theorem [5]. In Fig. S1, we plot the band-gap of the square lattice for increasing system sizes, calculated using periodic boundary conditions. We compare this to the true band-gap. Thus, we show that the band-gap is sufficiently converged.

## B. Design of the amorphous structure

The positions of the waveguides as shown in Figs. 1a and 1b are generated using a Metropolis Monte Carlo simulation of a liquid [15] with periodic boundary conditions, where the component atoms interact via an isotropic pair potential of the Yukawa form, $v(r) = v_0 e^{-(r-r_0)/l} / r$, where $r_0$=*14* $\mu m$ and $l$ =*2.8* $\mu m$. The length of the square box is *184.2 $\mu m$*. For our system, we take $k_B T/v_0$=*1.0* and the particle number density is $\rho = 2/\sqrt{3} r_0^2$, where $T$ is the temperature, $l$ is the screening length, and $k_B$ is Boltzmann's constant. At this temperature and density, the *N*-particle liquid is well above its freezing point.

## C. Waveguide fabrication

The amorphous photonic lattice employed here is fabricated by the laser direct-writing method in a fused silica sample [16]. We used a Ti:Sapphire laser system operating at a wavelength of *800nm*, a repetition rate of *100kHz* and a pulse length of *170fs*. A permanent change in the molecular structure of the material can be realized by tightly focusing ultrashort laser pulses into a transparent bulk material, causing nonlinear absorption. In fused silica, this induces a permanent increase in the refractive index with approximately the dimensions of the focus of the microscope objective focusing the writing beam. By moving the sample transversely with respect to the beam, a continuous modification of the refractive index is obtained, which creates a waveguide in the volume of the bulk silica. For the fabrication of our *2cm* long samples, the average power was adjusted to *32mW* and the writing velocity was set to *90mm/min*. The waveguides form a transverse refractive index profile of the form

$$n(x,y) = n_0 + \sum_{j=1}^{N} \Delta n_j e^{-\left((x-x_j)^2/\sigma_x^2 + (y-y_j)^2/\sigma_y^2\right)^3} \equiv n_0 + \Delta n(x,y)$$

which is invariant in the propagation direction *z*. The ambient refractive index of the fused silica is $n_0=1.45$, $N=200$ is the number of waveguides, $\Delta n_j$ and $(x_j, y_j)$ are the refractive index increase and position of the $j^{th}$ waveguide, respectively. The parameters $\sigma_x=1.5\mu m$ and $\sigma_y=1.5\mu m$ describe the transverse length and width of the waveguides, and $\Delta n(x,y)$, the deviation from the ambient refractive index, represents the potential in Eq. 1.